\documentclass[aps,prb,twocolumn,superscriptaddress,nofootinbib,longbibliography,floatfix]{revtex4-2}

\usepackage{amsmath,amssymb,bm}
\usepackage{graphicx}
\usepackage[colorlinks=true,citecolor=blue,linkcolor=blue,urlcolor=blue]{hyperref}

\newcommand{\LNO}{La$_3$Ni$_2$O$_7$}

\newcommand{\dxy}{d_{xy}}
\newcommand{\spm}{s_{\pm}}
\newcommand{\Gd}{G_{\rm DMFT}}

\begin{document}

\title{Correlation-renormalized spin-fluctuation pairing and the stabilization of $s_{\pm}$ superconductivity in pressurized La$_3$Ni$_2$O$_7$}

\author{Shuhong Tang}
\affiliation{Key Laboratory of Materials Physics, Institute of Solid State Physics, HFIPS, Chinese Academy of Sciences, Hefei 230031, China}
\affiliation{Science Island Branch of Graduate School, University of Science and Technology of China, Hefei 230026, China}

\author{Liang-Jian Zou}
\thanks{Corresponding author: \href{mailto:zou@theory.issp.ac.cn}{zou@theory.issp.ac.cn}.}
\affiliation{Key Laboratory of Materials Physics, Institute of Solid State Physics, HFIPS, Chinese Academy of Sciences, Hefei 230031, China}
\affiliation{Science Island Branch of Graduate School, University of Science and Technology of China, Hefei 230026, China}

\date{\today}

\begin{abstract}
The superconducting gap symmetry of pressurized La$_3$Ni$_2$O$_7$ remains unsettled because conventional weak-coupling calculations often place the system close to competing sign-changing $s$- and $d$-wave instabilities.  Using a realistic four-orbital Wannier Hamiltonian, we combine single-site two-orbital dynamical mean-field theory (DMFT) with a self-energy-renormalized random-phase approximation (RPA).  The central step is to replace the bare particle-hole bubble $G_0G_0$ of ordinary RPA by a $G_{\rm DMFT}G_{\rm DMFT}$ bubble, while keeping the same residual Slater--Kanamori interaction vertices.  In the bare RPA benchmark, the leading pairing eigenvalue belongs to the $B_{2g}$ $d_{xy}$ channel.  Once the DMFT self-energy is included, the hierarchy is reversed: the $A_{1g}$ sign-changing $s_{\pm}$ state becomes dominant, the $B_{1g}$ $d_{x^2-y^2}$ channel is subleading, and the original $B_{2g}$ instability is strongly suppressed.  Pocket-pair decomposition and orbital-resolved susceptibilities show that the reversal originates from orbital-selective renormalization of the $d_{3z^2-r^2}$ sector, which filters the $\gamma$-pocket scattering processes that stabilize $d_{xy}$ pairing in bare RPA while preserving distributed inter-pocket processes favorable to $s_{\pm}$ pairing.  As an independent two-particle validation, we further compute the static spin susceptibility using the dual Bethe--Salpeter equation with the local DMFT vertex.  The resulting susceptibility retains a broad finite-momentum magnetic response and is weak near $\Gamma$, strengthening the spin-fluctuation background for the correlation-stabilized $s_{\pm}$ state.  Our results demonstrate that strong correlations are not a secondary correction in La$_3$Ni$_2$O$_7$: an appropriate treatment of correlation-renormalized quasiparticles is essential for predicting the superconducting pairing symmetry.
\end{abstract}

\maketitle

\section{Introduction}

The discovery of superconducting signatures with transition temperatures approaching
$80$ K in pressurized \LNO\ has made the bilayer nickelate a central platform for testing
unconventional pairing in a multi-orbital correlated metal
\cite{Sun2023Nature,Zhang2024NatPhys,Wang2024PRX,Li2025NSR}.  The decisive open
question is the symmetry of the superconducting order parameter.  Because the low-energy
states are formed mainly by the Ni-$e_g$ orbitals $d_{x^2-y^2}$ and $d_{3z^2-r^2}$, the condensate may
realize a sign-changing $A_{1g}$ state, conventionally denoted $\spm$, or one of several
$d$-wave states rather than following a single-band analogy.  Recent experiments have made
this issue more concrete, but not yet settled it.  High-pressure Andreev-reflection
measurements reported multicomponent gap features, and a subsequent Blonder--Tinkham--
Klapwijk (BTK) analysis of \LNO$_{-\delta}$ found two distinct gaps consistent with an
$s$-like two-gap spectrum \cite{Liu2025SCPMA,Guo2025NatCommun}.  In strained bilayer
nickelate films, scanning tunneling microscopy/spectroscopy (STM/STS) and angle-resolved
photoemission spectroscopy (ARPES) have reported nodeless or anisotropic $s$-like gaps,
including flat-bottom U-shaped spectra with nearly vanishing low-energy density of states
\cite{Fan2025arxiv,Shen2025arxiv,Wang2026arxiv,Liang2026arxiv}.  Conversely, directional
point-contact spectroscopy on pressurized single crystals has been interpreted as evidence
for a predominant $d$-wave-like gap \cite{Cao2025arxiv}.  The experimental frontier
therefore points to a close competition between nodeless sign-changing $s$ wave and nodal
$d$ wave.

Most theoretical analyses start from a pressure-stabilized bilayer two-orbital Hamiltonian,
where the interlayer $d_{3z^2-r^2}$ hopping generates a pronounced bonding--antibonding structure and
the $\gamma$ sheet plays an essential role \cite{Luo2023PRL}.  Early weak-to-moderate
coupling spin-fluctuation calculations, including the random-phase approximation (RPA) and
functional renormalization group (FRG), found an $\spm$ state generated by repulsive
scattering between different Fermi-surface sheets
\cite{Liu2023PRL,Yang2023PRB,Zhang2024NatCommun}.  Related density functional theory
(DFT)-based RPA studies of pressure evolution and of the rare-earth 327 family also
emphasized an interlayer sign-changing $s$-wave tendency
\cite{Zhang2023PRBRareEarth,Jiang2025PRL}.  In parallel, strong-coupling and
nonperturbative treatments---bilayer $t$-$J$ models, density-matrix renormalization group
(DMRG) or ladder calculations, auxiliary-field or constrained-path quantum Monte Carlo
(QMC), cluster dynamical mean-field theory (cluster-DMFT), Gutzwiller-type approaches, and
recent dynamical-cluster QMC---have often identified interlayer magnetic exchange, especially
in the $d_{3z^2-r^2}$ sector, as the microscopic origin of extended $s$ or $\spm$ pairing
\cite{Qin2023PRB,Shen2023CPL,Sakakibara2024PRL,Oh2023PRB,Kaneko2024PRB,Ryee2024PRL,Tian2024PRB,Luo2024NPJ,Lu2024PRL,Qu2024PRL,Maier2026NPJ}.

Nevertheless, the weak-coupling spin-fluctuation literature also shows that the $\spm$
state is not automatically robust in ordinary RPA-type calculations.  Local-density approximation plus fluctuation-exchange (LDA+FLEX) calculations found strong competition
between $B_{2g}$ $\dxy$ and $A_{1g}$ $\spm$ channels near a spin-density-wave instability
\cite{Heier2024PRB}.  More strikingly, DFT--Wannier--RPA calculations based on an accurately
reproduced DFT band structure placed the $B_{2g}$ $\dxy$ state as the leading instability,
with $\spm$ becoming dominant only after a modest increase of the Ni-$e_g$ crystal-field
splitting \cite{Xia2025NatCommun}.  Realistic multi-orbital RPA studies including Ni $3d$
and O $2p$ states likewise found extended regions of their pressure- and interaction-
parameter phase diagrams in which a $d$-wave state, often of $d_{xy}$ character, preempts
sign-changing $s$-wave pairing \cite{XuEtAl2025Competition}.  FLEX calculations including
interlayer Coulomb interactions showed a possible transition from $\spm$ to a
$d_{x^2-y^2}$ state, while FRG studies of nonlocal repulsion found that interlayer repulsion
can promote $d_{x^2-y^2}$ pairing, although with a lower critical scale than the interlayer
$\spm$ state in realistic regimes \cite{Xi2025PRB,Zhan2026NPJ}.  Additional calculations
have further emphasized the sensitivity of the leading pairing channel to the $\gamma$
pocket, Hund coupling, crystal-field splitting, strain, and interaction range
\cite{Gao2025JPCM,Gao2025PhysicaC,Xiong2026arxiv,ZhangFilm2026PRB}.  The recurring message
is that ordinary weak-coupling RPA does not by itself yield a universally separated $\spm$
leading eigenvalue.  In many RPA or RPA-related calculations, $B_{2g}$ $\dxy$ is either the
leading state or occupies a substantial part of the theoretical phase diagram.  This provides
a sharp benchmark for testing whether the pairing symmetry predicted for \LNO\ is stable
once the correlated quasiparticle dynamics are treated more realistically.

This fragility motivates the central question of the present work.  Conventional RPA uses a
bare Green's function $G_0$ and bare Slater--Kanamori vertices, so the pairing kernel is
controlled mainly by unrenormalized Fermi-surface geometry and by proximity to a Stoner
instability.  This approximation is severe for \LNO.  ARPES, optical spectroscopy, and
first-principles many-body calculations indicate orbital-dependent mass renormalization,
reduced kinetic energy, and particularly strong correlation effects in the $d_{3z^2-r^2}$ sector
\cite{YangARPES2024NatCommun,LiuOptics2024NatCommun,Lechermann2023PRB,Christiansson2023PRL}.
Since the same $d_{3z^2-r^2}$ orbital and $\gamma$-sheet physics are central to both the $\spm$
mechanism and several competing $d$-wave channels, a prediction based only on the bare
Wannier quasiparticles can misidentify the leading symmetry.  The decisive comparison is
therefore not bare RPA versus a strong-coupling model, but ordinary RPA versus a
correlation-renormalized RPA pairing problem in which the same spin-fluctuation framework
is supplied with the correlated single-particle propagators.

In this work we perform this comparison for the four-orbital Wannier model of \LNO.  We use
a single-site DMFT self-energy to replace the bare RPA bubble by a $\Gd\Gd$ bubble and then
solve the symmetry-resolved static pairing equation.  This correlation-renormalized RPA
calculation converts the bare-RPA $B_{2g}$ $d_{xy}$ instability into a robust $A_{1g}$
$\spm$ leading state, with the reversal traced to the selective suppression of
$\gamma$-pocket scattering and the survival of distributed inter-pocket processes favorable
to sign-changing $s$ wave.  A static spin susceptibility calculated with the dual
Bethe--Salpeter equation (DBSE) further confirms that the finite-momentum magnetic response
underlying the $\spm$ analysis survives after local DMFT vertex corrections are included.
The remainder of the paper presents the model and methods, the numerical results, and the
implications for correlation-controlled pairing symmetry in \LNO.

\section{Computational Methods}
\label{sec:methods}

\subsection{Four-orbital Wannier Hamiltonian}
\label{sec:model_methods}

We use the four-orbital Wannier Hamiltonian constructed by Xia \textit{et al.} for
\LNO~\cite{Xia2025NatCommun}.  The real-space hopping parameters are taken from
Supplementary Data~1 of Ref.~\onlinecite{Xia2025NatCommun} and parsed in the
\textsc{wannier90} \texttt{HR} format~\cite{Pizzi2020Wannier90}.  The orbital basis is
fixed as
\begin{equation}
  (1z,1x,2z,2x),
  \label{eq:orbital_basis_methods}
\end{equation}
where $z\equiv d_{3z^2-r^2}$, $x\equiv d_{x^2-y^2}$, and 1 and 2 denote the two NiO$_2$ layers.  The Bloch Hamiltonian is obtained from
\begin{equation}
  H_{ab}(\mathbf k)=\sum_{\mathbf R}H_{ab}(\mathbf R)e^{i\mathbf k\cdot\mathbf R},
  \label{eq:hk_hr_methods}
\end{equation}
with $a,b$ running over the four Wannier orbitals.  We work at $T=116$ K
($\beta\simeq100~{\rm eV}^{-1}$) and fix the total filling to $n=3$ electrons per bilayer
unit cell.  The Fermi surface used in the gap equation is discretized into about
$1.3\times10^3$ patches obtained from a dense two-dimensional momentum mesh; for each patch
we store the band index, orbital eigenvector, Fermi velocity, and line-element weight.

\subsection{Single-site two-orbital DMFT self-energy}
\label{sec:dmft_methods}

Local correlations are treated by single-site two-orbital DMFT~\cite{Georges1996DMFT},
implemented with the Toolbox for Research on Interacting Quantum Systems (\textsc{TRIQS}) framework and its continuous-time quantum Monte Carlo
impurity solver~\cite{Parcollet2015TRIQS,Seth2016TRIQSCTHYB}.  Since the two layers are
symmetry equivalent, a single two-orbital impurity problem is solved for the two $e_g$
orbitals, denoted by $z$ and $x$ in the equations below, and the resulting self-energy is
embedded into the bilayer lattice as
\begin{equation}
  \Sigma_{\rm lat}(i\omega_n)=
  \operatorname{diag}\left[
  \Sigma_{z}(i\omega_n),\Sigma_{x}(i\omega_n),
  \Sigma_{z}(i\omega_n),\Sigma_{x}(i\omega_n)\right].
  \label{eq:sigma_embedding_methods}
\end{equation}
The impurity interaction is the density-density Kanamori form with
$U_{\rm DMFT}=4.0$ eV and $J_{\rm DMFT}=0.5$ eV.  A Held/Kanamori double-counting correction
is applied~\cite{Held2007AdvPhys}.  The correlated lattice Green's function is then
\begin{equation}
  G_{\rm DMFT}(\mathbf k,i\omega_n)=
  \left[(i\omega_n+\mu)\mathbf 1-H(\mathbf k)-
  \Sigma_{\rm lat}^{\rm eff}(i\omega_n)\right]^{-1},
  \label{eq:gdmft_methods}
\end{equation}
where $\Sigma_{\rm lat}^{\rm eff}$ denotes the double-counting-corrected self-energy.  The
quasiparticle weights estimated from the low-frequency self-energy are approximately
$Z_z\simeq0.47$ and $Z_x\simeq0.63$, showing stronger renormalization in the
$d_{3z^2-r^2}$ orbital.  This orbital selectivity is the central single-particle input to
our self-energy-renormalized RPA calculation.

\subsection{DMFT-dressed RPA susceptibility}
\label{sec:gdmft_rpa_methods}

The central approximation of this work is a self-energy-renormalized RPA.  Compared with
ordinary multi-orbital RPA, the bare particle-hole bubble $G_0G_0$ is replaced by a bubble
constructed from the DMFT Green's function in Eq.~\eqref{eq:gdmft_methods}, while the onsite
Slater--Kanamori RPA vertices are kept unchanged.  This isolates the effect of the
orbital-dependent DMFT self-energy on the spin-fluctuation pairing problem.

The static irreducible susceptibility is evaluated as
\begin{equation}
\begin{split}
  \chi^{0,\Sigma}_{l_1l_2l_3l_4}(\mathbf q)
  =-&\frac{T}{N_k}\sum_{\mathbf k,n}
  G_{\rm DMFT}^{l_3l_1}(\mathbf k+\mathbf q,i\omega_n)\\
  &\times G_{\rm DMFT}^{l_2l_4}(\mathbf k,i\omega_n),
\end{split}
  \label{eq:chi0_gdmft_methods}
\end{equation}
where the orbital indices refer to the basis in Eq.~\eqref{eq:orbital_basis_methods}.  The
spin and charge RPA susceptibilities are obtained by matrix inversion in composite orbital
space,
\begin{align}
  \chi_s^{\Sigma,{\rm RPA}}(\mathbf q)
  &=\left[1-\chi^{0,\Sigma}(\mathbf q)U^s\right]^{-1}\chi^{0,\Sigma}(\mathbf q),
  \nonumber\\
  \chi_c^{\Sigma,{\rm RPA}}(\mathbf q)
  &=\left[1+\chi^{0,\Sigma}(\mathbf q)U^c\right]^{-1}\chi^{0,\Sigma}(\mathbf q),
  \label{eq:rpa_chi_methods}
\end{align}
with the standard spin and charge Slater--Kanamori matrices $U^s$ and $U^c$
\cite{Graser2009NJP,Graser2010PRB,Liu2023PRL,Xia2025NatCommun}.  The residual RPA
interactions are denoted $U_{\rm RPA}$ and $J_{\rm RPA}$; throughout the comparison we take
$J_{\rm RPA}=0.15U_{\rm RPA}$, with $U'=U-2J$ and $J'=J$.  For selected analyses we tune
$U_{\rm RPA}$ so that the Stoner factor $\eta_S$ is approximately 0.95 in both the bare and
DMFT-dressed calculations, enabling a comparison at the same proximity to magnetic order.

\subsection{Pairing vertex and symmetry-resolved gap equation}
\label{sec:pairing_kernel_methods}

The static singlet pairing vertex is constructed from the RPA susceptibilities using the
standard multiorbital spin-fluctuation expression,
\begin{align}
  \Gamma^{\Sigma,{\rm RPA}}(\mathbf q)
  =&\ \frac32 U^s\chi_s^{\Sigma,{\rm RPA}}(\mathbf q)U^s
  -\frac12 U^c\chi_c^{\Sigma,{\rm RPA}}(\mathbf q)U^c
  \nonumber\\
  &+\frac12\left(U^s+U^c\right),
  \label{eq:pairing_vertex_methods}
\end{align}
where all products are matrix products in four-orbital composite-index space.  The ordinary
RPA reference is obtained from the same equations with $G_0$ replacing $G_{\rm DMFT}$ in the
bubble.

The orbital-space pairing vertex is projected onto the Fermi-surface patches using the
Wannier eigenvectors and singlet symmetrization.  This gives a patch kernel $K_{ij}$, and the
pairing strengths follow from
\begin{equation}
  \lambda\Delta_i=\sum_jK_{ij}\Delta_j .
  \label{eq:linear_gap_patch_methods}
\end{equation}
With our sign convention, a positive eigenvalue corresponds to an attractive instability in
the corresponding gap channel.  To resolve the symmetry, $K$ is projected onto the even-parity
irreducible representations of the in-plane $C_{4v}$ group,
\begin{equation}
  P_\Gamma K P_\Gamma \Delta_\Gamma=\lambda_\Gamma\Delta_\Gamma,
  \qquad
  \Gamma=A_{1g},B_{1g},B_{2g},A_{2g}.
  \label{eq:sym_gap_methods}
\end{equation}
The $A_{1g}$ solution is identified as $s_\pm$ when the gap changes sign between the
relevant Fermi-surface sheets, while the $B_{1g}$ and $B_{2g}$ sectors correspond to
$d_{x^2-y^2}$- and $d_{xy}$-like gap structures.  For the pocket-resolved analysis in
Sec.~\ref{sec:symmetry_reversal_mechanism}, the eigenvalue is decomposed into contributions
from the three Fermi-surface sheets $\alpha$, $\beta$, and $\gamma$ by evaluating the
bilinear form of $K$ within each pair of pockets.

\subsection{DBSE spin susceptibility validation}
\label{sec:dbse_methods}

As a complementary vertex-level validation of the correlated magnetic spectrum, we compute
$\chi_{S_zS_z}(\mathbf q)$ with the dual Bethe--Salpeter equation (DBSE) formulation
implemented in the two-particle response-function (TPRF) library of \textsc{TRIQS}~\cite{TriqsTPRF,VanLoon2024DBSE}.  The calculation uses
the same four-orbital Hamiltonian, local interaction parameters, and DMFT framework as the
self-energy-renormalized RPA analysis.  Starting from the converged impurity problem, the
local two-particle quantities are used to construct the reducible particle-hole vertex,
which is embedded into the lattice DBSE to obtain the generalized susceptibility
$\chi^{\rm DBSE}_{l_1l_2l_3l_4}(\mathbf q,\Omega)$.  The physical spin response is obtained
by contracting the spin components to $\chi^{\rm DBSE}_{S_zS_z}(\mathbf q,\Omega)$.

The DBSE validation reported below is performed in the static limit, $\Omega=0$, for the
$q_z=0$ plane.  We use $\beta=25~{\rm eV}^{-1}$ and retain $N_{\rm wf}=10$ fermionic
Matsubara frequencies in the particle-hole vertex.  This calculation directly probes
whether the finite-momentum spin-fluctuation structure that drives the
$G_{\rm DMFT}G_{\rm DMFT}$-RPA pairing hierarchy is preserved after the local DMFT
vertex correction is included.

\section{Numerical result}
\label{sec:results}

\subsection{Orbital-selective correlated electronic structure}
\label{sec:dmft_electronic_structure}

\begin{figure*}[!t]
  \centering
  \includegraphics[width=\textwidth]{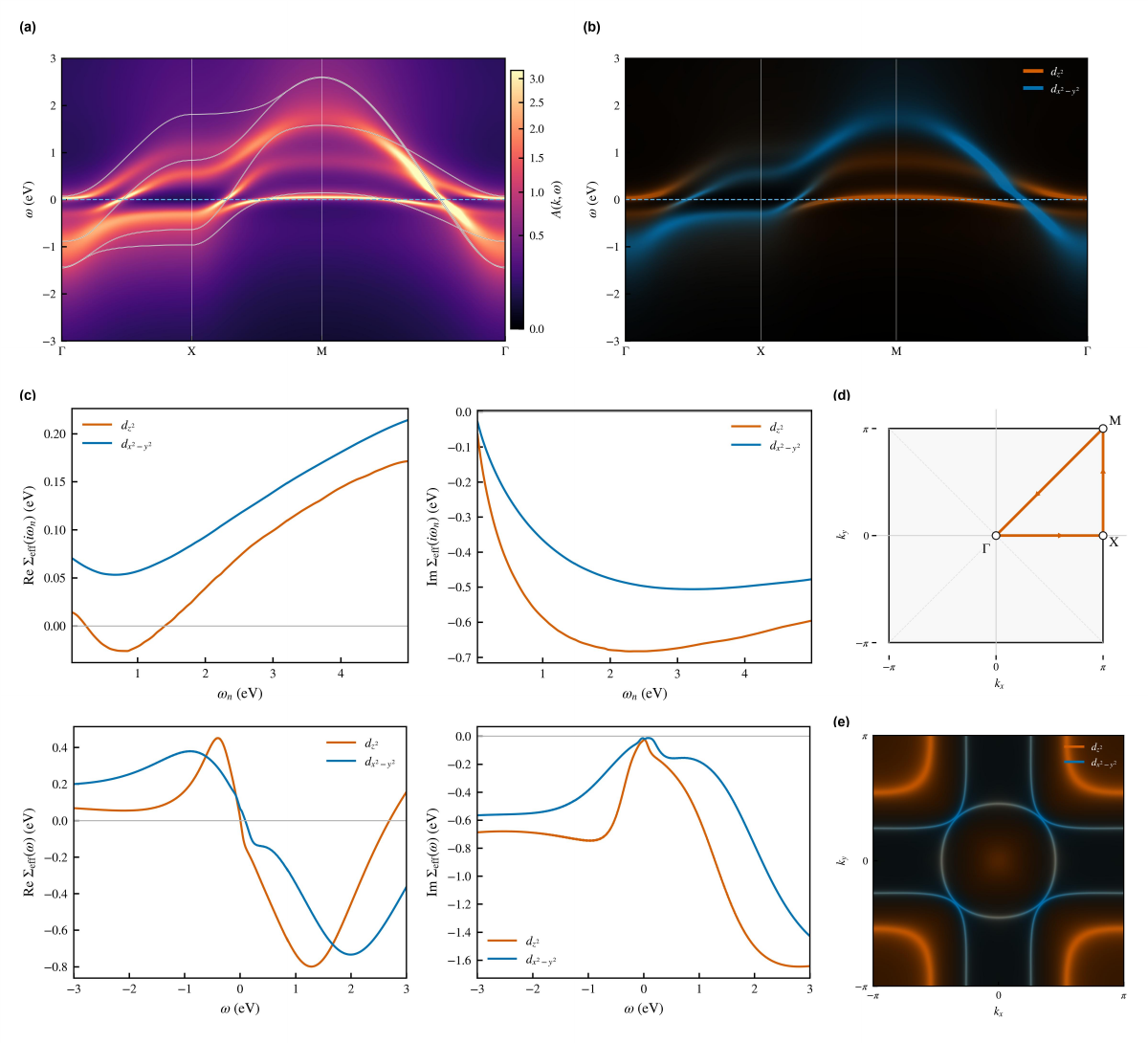}
  \caption{(Colour online) DMFT-renormalized electronic structure of the four-orbital Wannier model.
  (a) Total spectral function $A(\mathbf k,\omega)$ along the high-symmetry path
  $\Gamma$--$X$--$M$--$\Gamma$.  The gray curves indicate the underlying Wannier bands,
  while the color scale shows the correlated spectral weight.  (b) Orbital-resolved
  spectral intensity, separating the $d_{3z^2-r^2}$ and $d_{x^2-y^2}$ contributions.
  (c) Orbital-resolved DMFT self-energy on the Matsubara axis and in a real-frequency
  representation.  The steeper low-frequency slope of the $d_{3z^2-r^2}$ self-energy gives
  a smaller quasiparticle weight than for $d_{x^2-y^2}$.  (d) Brillouin-zone convention and
  high-symmetry path used in panels (a) and (b).  (e) Orbital-resolved zero-energy spectral
  weight in the Brillouin zone.  The orbital differentiation visible in panels (b), (c), and
  (e) is the single-particle input responsible for the correlation-induced reweighting of
  the RPA susceptibility and pairing kernel.}
  \label{fig:dmft_electronic_structure}
\end{figure*}

Before discussing the superconducting eigenvalues, we first characterize the correlated
normal state that enters the DMFT-dressed RPA calculation.  This step is essential because
our approximation does not alter the residual RPA interaction vertices; all correlation
corrections to the spin-fluctuation kernel originate from the replacement of the bare
propagator by $G_{\rm DMFT}$ in the particle-hole bubble.  The single-particle spectra in
Fig.~\ref{fig:dmft_electronic_structure} therefore provide the physical link between the
Wannier Hamiltonian of Ref.~\onlinecite{Xia2025NatCommun} and the pairing hierarchy
analyzed below.

Fig.~\ref{fig:dmft_electronic_structure}(a) shows that the low-energy bands of the
Xia \textit{et al.} Wannier model remain well resolved after the single-site DMFT self-energy
is included.  The coherent spectral weight still follows the principal bilayer $e_g$ bands
through the Fermi level, so the Fermi-surface sheet assignment used in the patch construction
remains meaningful.  At the same time, the spectral peaks are no longer described by sharp
bare bands: away from the Fermi level the intensity is broadened and redistributed, signaling
that the particle-hole response entering RPA must be viewed as a spectral-function-weighted
nesting problem rather than as a purely geometric nesting problem of $G_0$.

The orbital decomposition in Fig.~\ref{fig:dmft_electronic_structure}(b) identifies the
origin of this reweighting.  The $d_{3z^2-r^2}$ component contributes strongly to the
near-Fermi-level bilayer-derived spectral weight, whereas the $d_{x^2-y^2}$ component forms
more dispersive features along the same high-symmetry path.  This distinction is central for
La$_3$Ni$_2$O$_7$: weak-coupling theories often associate the sign-changing $A_{1g}$ state
with scattering processes involving the bilayer $d_{3z^2-r^2}$ sector, but the same orbital
is also the one most affected by correlations in both many-body calculations and
spectroscopy~\cite{YangARPES2024NatCommun,LiuOptics2024NatCommun,Lechermann2023PRB,Christiansson2023PRL}.
The present calculation treats this correlation effect explicitly at the propagator level.

The self-energy in Fig.~\ref{fig:dmft_electronic_structure}(c) quantifies the orbital
selectivity.  From the low-frequency Matsubara slope we obtain
quasiparticle weights of approximately $Z_z\simeq0.47$ and $Z_x\simeq0.63$, with the
$d_{3z^2-r^2}$ orbital being more strongly renormalized.  Thus the correlated
Green's function entering the susceptibility contains a smaller coherent weight and a larger
damping in the $d_{3z^2-r^2}$ component than in the $d_{x^2-y^2}$ component.  This result is not used as an empirical rescaling factor; rather, the full
Matsubara-axis self-energy is retained in the bubble
$\chi^{0,\Sigma}\sim -T\sum_k G_{\rm DMFT}(k+q)G_{\rm DMFT}(k)$.  Consequently, both the
reduction of quasiparticle weight and the finite lifetime broadening are included before the
RPA ladder is summed.

The momentum-space consequence is summarized in Fig.~\ref{fig:dmft_electronic_structure}(e),
which displays the orbital-resolved zero-energy spectral weight.  The Fermi surface remains
multi-sheeted, but the coherent spectral weight is strongly orbital dependent.  Since the
spin-fluctuation pairing vertex is obtained from the matrix susceptibility
$\chi_s^{\Sigma,{\rm RPA}}(\mathbf q)$, this orbital texture determines which scattering
processes are amplified and which are filtered out by the DMFT self-energy.  In a bare RPA
calculation, two Fermi-surface segments with favorable nesting contribute according to their
band geometry and orbital coherence factors alone.  In the present $G_{\rm DMFT}$-RPA
calculation, the same processes are additionally weighted by the orbital-dependent spectral
coherence and damping shown in Fig.~\ref{fig:dmft_electronic_structure}.  This is the
mechanism by which the correlated normal state can change the relative strength of the
$A_{1g}$, $B_{1g}$, and $B_{2g}$ pairing channels without changing the formal RPA vertex.

The main conclusion of this subsection is therefore twofold.  First, the DMFT solution does
not invalidate the low-energy four-orbital Wannier description: the bilayer $e_g$ Fermi
surfaces remain sufficiently coherent to support a controlled comparison with the ordinary
RPA benchmark.  Second, the quasiparticles are strongly and selectively renormalized, with $Z_z<Z_x$, i.e., the $d_{3z^2-r^2}$ quasiparticle weight is smaller than the $d_{x^2-y^2}$ quasiparticle weight.  The following subsections show that this orbital-selective propagator produces a
qualitative reversal of the RPA pairing hierarchy and then trace that reversal to the
susceptibility and pocket-pair decomposition of the pairing kernel.

\subsection{Correlation-induced reversal of the pairing hierarchy}
\label{sec:pairing_hierarchy}

We now turn to the central result of this work: the evolution of the leading
spin-fluctuation pairing channels when the ordinary RPA bubble is replaced by its
DMFT-renormalized counterpart.  The comparison is deliberately constructed so that the
bare and correlated calculations use the same four-orbital Wannier Hamiltonian, the same
Fermi-surface patches, and the same Slater--Kanamori RPA vertex matrices.  The only formal
change is
\begin{equation}
  G_0G_0 \longrightarrow G_{\rm DMFT}G_{\rm DMFT}
  \label{eq:g0_to_gdmft_results}
\end{equation}
in the irreducible particle-hole susceptibility entering the RPA ladder.  Consequently, any
change in the pairing hierarchy can be traced to the orbital- and frequency-dependent DMFT
self-energy discussed in Sec.~\ref{sec:dmft_electronic_structure}, rather than to a change
of the residual interaction parameters.

\begin{figure*}[!t]
  \centering
  \includegraphics[width=\textwidth]{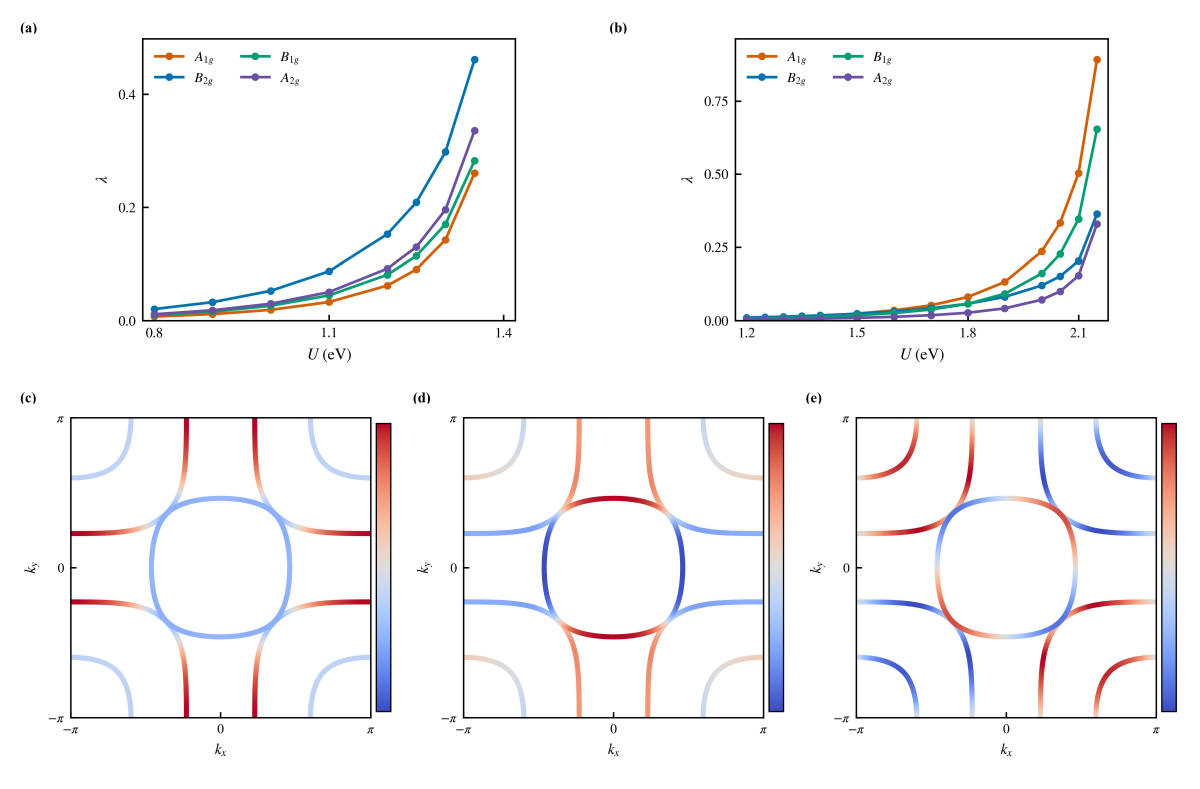}
  \caption{(Colour online) Correlation-induced reversal of the leading pairing symmetry.  (a) Symmetry-
  resolved pairing eigenvalues $\lambda_\Gamma$ obtained in ordinary RPA from the bare
  Wannier Green's function.  The $B_{2g}$ channel is the leading instability over the
  interaction range shown.  (b) Corresponding eigenvalues after replacing the irreducible
  bubble by $G_{\rm DMFT}G_{\rm DMFT}$.  The overall growth is shifted to larger residual
  $U_{\rm RPA}$ because the DMFT self-energy suppresses the coherent particle-hole
  response, but the leading symmetry is inverted to $A_{1g}$.  (c)--(e) Representative gap
  eigenvectors on the Fermi-surface sheets in the $A_{1g}$, $B_{1g}$, and $B_{2g}$ sectors.
  The $A_{1g}$ state is a sign-changing $s_\pm$ state, whereas $B_{1g}$ and $B_{2g}$ have
  $d_{x^2-y^2}$- and $d_{xy}$-like sign structures, respectively.}
  \label{fig:pairing_hierarchy}
\end{figure*}

Fig.~\ref{fig:pairing_hierarchy}(a) shows the conventional RPA benchmark.  As expected
for a repulsive spin-fluctuation pairing problem, all eigenvalues increase rapidly as
$U_{\rm RPA}$ approaches the magnetic Stoner regime.  The crucial feature is the symmetry
ordering.  In the bare calculation the $B_{2g}$ channel is the leading instability throughout
the range displayed, while the $A_{1g}$ and $B_{1g}$ channels remain lower in eigenvalue.
Near the upper end of the bare-RPA range, the $B_{2g}$ eigenvalue is visibly separated from
$A_{1g}$.  This behavior is consistent with realistic weak-coupling studies in which the
leading state of bilayer nickelate models depends sensitively on crystal-field splitting,
Fermi-surface shape, and interaction parameters: ordinary RPA does not generically produce
a well-separated $s_\pm$ eigenvalue, but often yields a close competition with prominent
$d$-wave channels, especially $B_{2g}$ $d_{xy}$
\cite{Heier2024PRB,Xia2025NatCommun,XuEtAl2025Competition}.

The inclusion of the DMFT self-energy qualitatively changes this hierarchy.  In
Fig.~\ref{fig:pairing_hierarchy}(b), the absolute growth of the pairing eigenvalues is
shifted to larger $U_{\rm RPA}$.  This shift is expected: replacing $G_0$ by the correlated
Green's function reduces the coherent spectral weight in the particle-hole bubble and moves
the RPA system farther from the Stoner boundary at the same residual interaction.  The more
important observation is that the symmetry ordering is reversed.  The $A_{1g}$ eigenvalue
becomes the dominant one over the correlated-RPA interaction range, while the $B_{2g}$
channel that led the bare calculation is pushed below both $A_{1g}$ and, at large
$U_{\rm RPA}$, the $B_{1g}$ channel.  The DMFT correction therefore cannot be represented
as a uniform multiplicative renormalization of all eigenvalues.  It changes the internal
structure of the pairing kernel and converts the leading instability from $B_{2g}$ $d_{xy}$
to $A_{1g}$ $s_\pm$.

The representative gap eigenvectors in Figs.~\ref{fig:pairing_hierarchy}(c)--(e) identify
the competing states.  The $A_{1g}$ solution preserves the point-group symmetry of the
lattice but changes sign between Fermi-surface regions, and is therefore an $s_\pm$ state
rather than a conventional sign-preserving $s$ wave.  Because this sign reversal is not
enforced by a change of irreducible representation, the state need not contain symmetry-
protected nodes and is naturally compatible with the nodeless and U-shaped spectra reported
in recent tunneling and Andreev-reflection experiments on bilayer nickelates
\cite{Guo2025NatCommun,Fan2025arxiv,Wang2026arxiv,Liang2026arxiv}.  By contrast, the
$B_{1g}$ state has the sign structure of $d_{x^2-y^2}$ pairing, while the $B_{2g}$ state has
$d_{xy}$ character.  These $d$-wave states remain legitimate eigenmodes of the repulsive
spin-fluctuation kernel, but after the DMFT self-energy is included they are no longer the
leading instability.

This symmetry reversal shows that the $s_\pm$ state is not merely an accidental weak-coupling solution of a particular bare
Fermi surface.  Instead, orbital-selective correlations filter the particle-hole processes
that enter the RPA ladder, penalizing the bare-RPA $B_{2g}$ channel more strongly than the
sign-changing $A_{1g}$ channel.  We next analyze this filtering mechanism by decomposing
the pairing eigenvalues into Fermi-surface pocket pairs and by comparing the bare susceptibilities and
DMFT-renormalized susceptibilities.

\begin{figure*}[!t]
\centering
\includegraphics[width=0.94\textwidth]{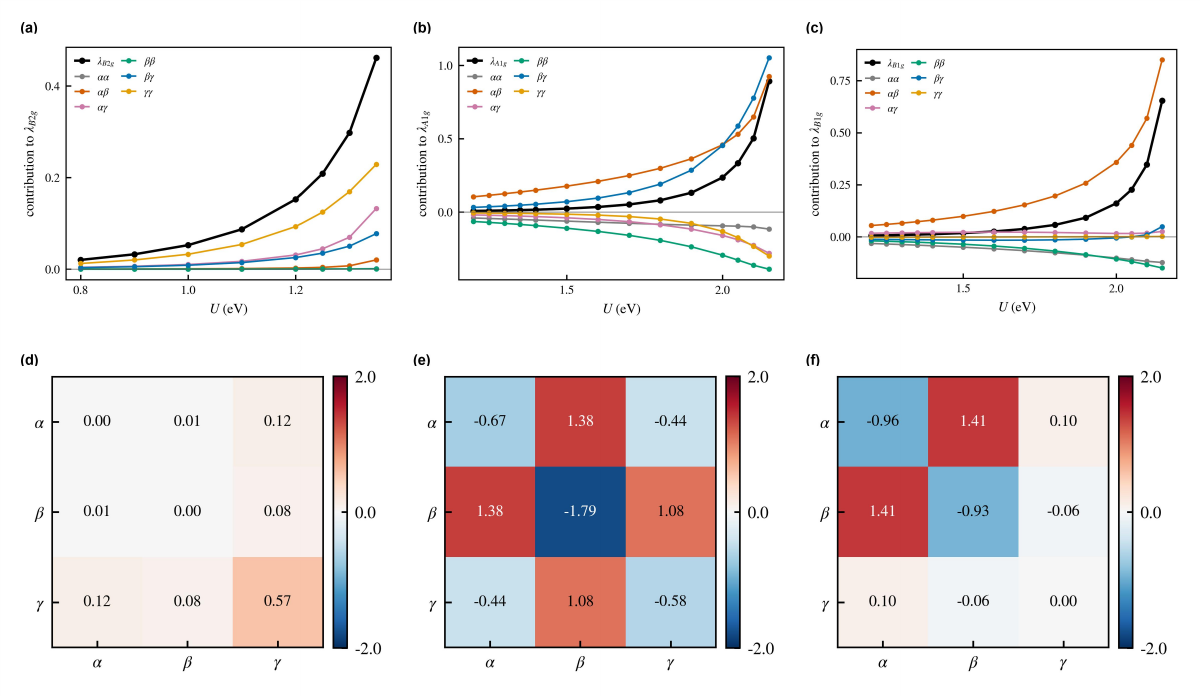}
\caption{(Colour online) Pocket-pair decomposition of the pairing eigenvalues.  (a) Contributions to the
ordinary-RPA $B_{2g}$ eigenvalue as a function of $U_{\rm RPA}$.  (b) and (c) Corresponding
decompositions for the DMFT-renormalized $A_{1g}$ and $B_{1g}$ channels.  The black curve
is the total eigenvalue, while the colored curves denote individual pocket pairs.  (d)--(f)
Normalized pocket-pair matrices $\lambda_\Gamma^{pp'}/\lambda_\Gamma$ at fixed Stoner
factor $\eta_S\simeq0.95$.  The bare $B_{2g}$ instability is dominated by the
$\gamma\gamma$ block, whereas the correlated $A_{1g}$ state is stabilized by large
inter-pocket $\alpha\beta$ and $\beta\gamma$ contributions.}
\label{fig:pocket_decomposition}
\end{figure*}

\subsection{Origin of the symmetry reversal: pocket selectivity and susceptibility reweighting}
\label{sec:symmetry_reversal_mechanism}

The reversal in Fig.~\ref{fig:pairing_hierarchy} is not a consequence of comparing two
calculations at different distances from a magnetic instability.  It reflects a change in the
internal structure of the pairing kernel.  To expose this structure we first decompose the
pairing eigenvalues into contributions from the three Fermi-surface sheets, and then compare
the irreducible magnetic susceptibilities that enter the RPA ladder.  This order emphasizes
which scattering processes select the gap symmetry before tracing their change back to the
DMFT self-energy.

For a normalized gap eigenvector $\Delta_\Gamma$, we write the pocket-pair contribution as
\begin{equation}
  \lambda_\Gamma^{pp'}=
  \sum_{i\in p}\sum_{j\in p'} w_i\,
  \Delta_{\Gamma,i}^{*}K_{ij}\Delta_{\Gamma,j},
  \label{eq:pocket_decomposition_results}
\end{equation}
where $p,p'\in\{\alpha,\beta,\gamma\}$ label Fermi-surface sheets and $K_{ij}$ is the
symmetrized patch kernel.  The sum over all $p,p'$ reproduces the full eigenvalue.  Positive
entries identify scattering processes that are used constructively by the gap form factor,
whereas negative entries are pair-breaking for that symmetry.  In the fixed-Stoner comparison
below, $U_{\rm RPA}$ is chosen separately in the bare and DMFT-dressed calculations so that
the Stoner factor $\eta_S\simeq0.95$, close to the RPA magnetic upper bound.

Fig.~\ref{fig:pocket_decomposition}(a) shows that the bare-RPA $B_{2g}$ state is a highly
concentrated instability.  Its total eigenvalue closely follows the $\gamma\gamma$
contribution, with smaller positive assistance from $\alpha\gamma$ and $\beta\gamma$
scattering and almost no support from $\alpha\alpha$, $\beta\beta$, or $\alpha\beta$ blocks.
The fixed-Stoner matrix in Fig.~\ref{fig:pocket_decomposition}(d) makes the same point
without reference to the absolute value of $U_{\rm RPA}$: more than half of the normalized
$B_{2g}$ eigenvalue comes from the $\gamma\gamma$ sector.  Thus ordinary RPA selects the
$d_{xy}$ form factor through a relatively narrow set of $\gamma$-centered processes.

The DMFT-renormalized $A_{1g}$ state has a qualitatively different structure.  In
Fig.~\ref{fig:pocket_decomposition}(b), the large positive terms are distributed over
inter-pocket channels, especially $\alpha\beta$ and $\beta\gamma$, while several diagonal
blocks are negative.  At fixed Stoner factor, Fig.~\ref{fig:pocket_decomposition}(e), the
positive off-diagonal matrix elements are partially compensated by pair-breaking intra-pocket
terms such as $\beta\beta$.  This is the characteristic fingerprint of an $s_\pm$ state: the
gap changes sign between pockets connected by strong repulsive spin fluctuations, thereby
turning inter-pocket repulsion into a positive pairing contribution.  Because the attraction
comes from a network of inter-pocket processes rather than from a single $\gamma\gamma$
block, the $A_{1g}$ state is more robust against the orbital-selective loss of coherent
$\gamma$-sheet weight.

The correlated $B_{1g}$ channel, shown in Figs.~\ref{fig:pocket_decomposition}(c) and
\ref{fig:pocket_decomposition}(f), explains why $d_{x^2-y^2}$ becomes the closest competitor
after the DMFT correction.  It also gains strongly from $\alpha\beta$ scattering.  However,
it obtains little additional support from the $\beta\gamma$ and $\gamma\gamma$ sectors, and
it pays sizable negative contributions in the $\alpha\alpha$ and $\beta\beta$ blocks.  The
$B_{1g}$ form factor therefore captures one important inter-pocket process but lacks the
broader cooperative structure of the $A_{1g}$ solution.  This is why the correlated RPA
calculation produces a subleading $B_{1g}$ state rather than restoring the bare-RPA $B_{2g}$
leader.

\begin{figure*}[!t]
\centering
\includegraphics[width=0.82\textwidth]{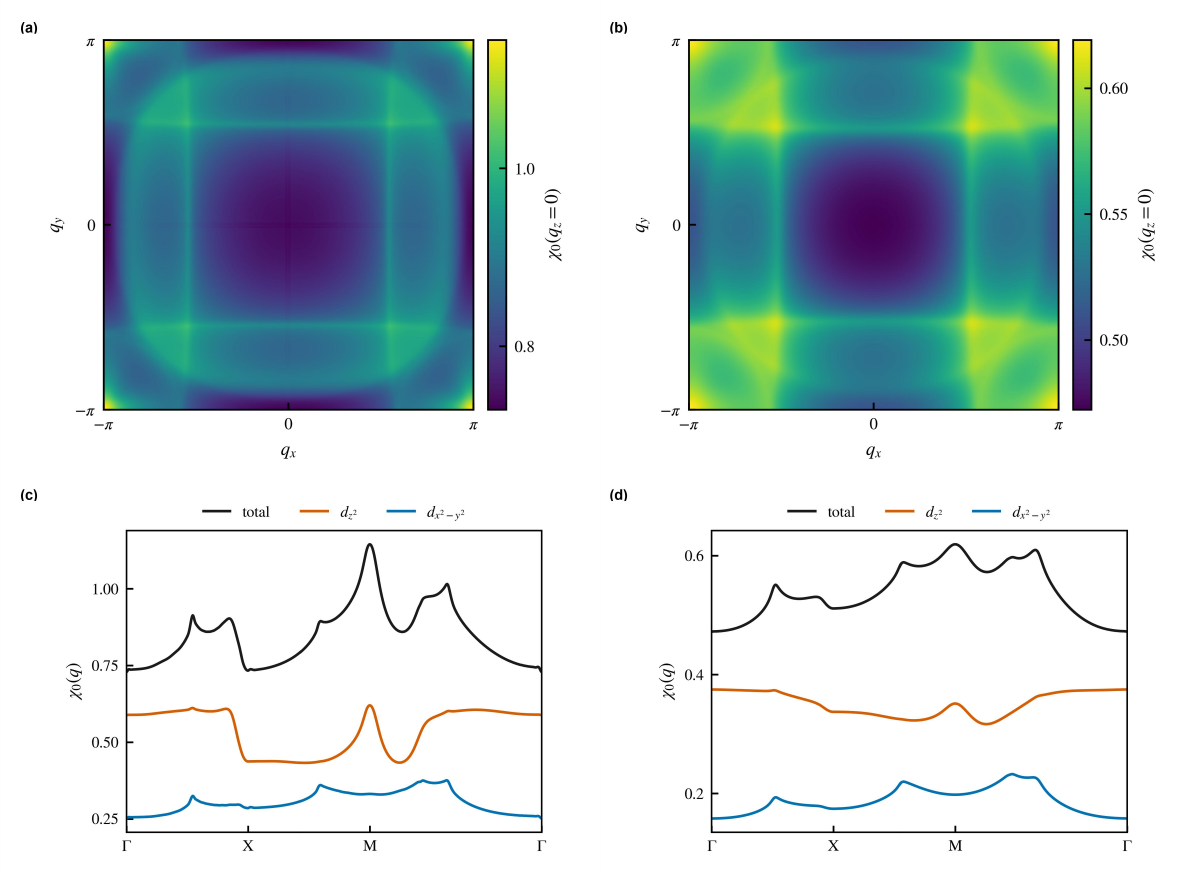}
\caption{(Colour online) Self-energy reweighting of the static irreducible magnetic susceptibility entering
  the RPA ladder.  (a) Momentum dependence of the bare bubble $\chi^0(\mathbf q,q_z=0)$
  obtained from $G_0G_0$.  (b) DMFT-dressed bubble $\chi^{0,\Sigma}(\mathbf q,q_z=0)$
  obtained from $G_{\rm DMFT}G_{\rm DMFT}$.  (c),(d) Corresponding total and
  orbital-resolved cuts along $\Gamma$--$X$--$M$--$\Gamma$.  The DMFT self-energy reduces
  the overall scale of the bubble and suppresses the $d_{3z^2-r^2}$ contribution more
  strongly than a uniform rescaling would.}
\label{fig:chi0_reweighting}
\end{figure*}

The susceptibility data in Fig.~\ref{fig:chi0_reweighting} reveal why the $\gamma$-centered
$B_{2g}$ route is selectively penalized.  We compare the static irreducible magnetic bubble
constructed from the bare Wannier propagator, $\chi^0$, with the DMFT-dressed bubble,
$\chi^{0,\Sigma}$.  These are the particle-hole kernels that are amplified by the RPA spin
ladder; they are not yet the final RPA susceptibilities.

The correlated bubble is smaller in magnitude throughout the Brillouin zone, which explains
why the eigenvalues in Fig.~\ref{fig:pairing_hierarchy}(b) grow rapidly only at larger
$U_{\rm RPA}$ than in the bare calculation.  This is the expected Stoner-scale effect of
using $G_{\rm DMFT}$: quasiparticle weight is reduced and part of the spectral weight is
shifted into incoherent states.  More importantly, the reduction is orbital and momentum
selective.  The line cuts in Figs.~\ref{fig:chi0_reweighting}(c) and
\ref{fig:chi0_reweighting}(d) show that the $d_{3z^2-r^2}$ contribution, which is prominent
in the bare response and tied to the bilayer $\gamma$-sheet physics, is suppressed in a way
consistent with $Z_z<Z_x$ obtained from DMFT.

Combining Figs.~\ref{fig:pocket_decomposition} and \ref{fig:chi0_reweighting} gives the
mechanism of the symmetry reversal.  Bare RPA favors $B_{2g}$ because it overemphasizes a
sharp, coherent, $\gamma$-rich particle-hole response.  The DMFT self-energy reduces and
broadens precisely that response, thereby weakening the $\gamma\gamma$ contribution that
made the $d_{xy}$ state dominant.  The sign-changing $A_{1g}$ state instead uses a more
distributed set of residual inter-pocket processes, primarily $\alpha\beta$ and
$\beta\gamma$, and therefore survives the correlation-induced filtering more efficiently.
The fixed-Stoner decomposition demonstrates that this is a genuine reorganization of the
pairing kernel, not merely the result of lowering the overall susceptibility amplitude.

\subsection{DBSE validation of the correlated magnetic fluctuation spectrum}
\label{sec:dbse_validation}

The analysis above attributes the $B_{2g}\rightarrow A_{1g}$ reversal to the way in which
the DMFT self-energy reweights the spin-fluctuation kernel.  To strengthen this result
beyond the self-energy-dressed bubble level, we compute the static spin susceptibility using
the dual Bethe--Salpeter equation.  This DBSE calculation includes the local DMFT
particle-hole vertex and therefore provides an independent two-particle test of the magnetic
fluctuation spectrum that underlies the $G_{\rm DMFT}G_{\rm DMFT}$-RPA pairing analysis.

\begin{figure*}[!t]
\centering
\includegraphics[width=0.78\textwidth]{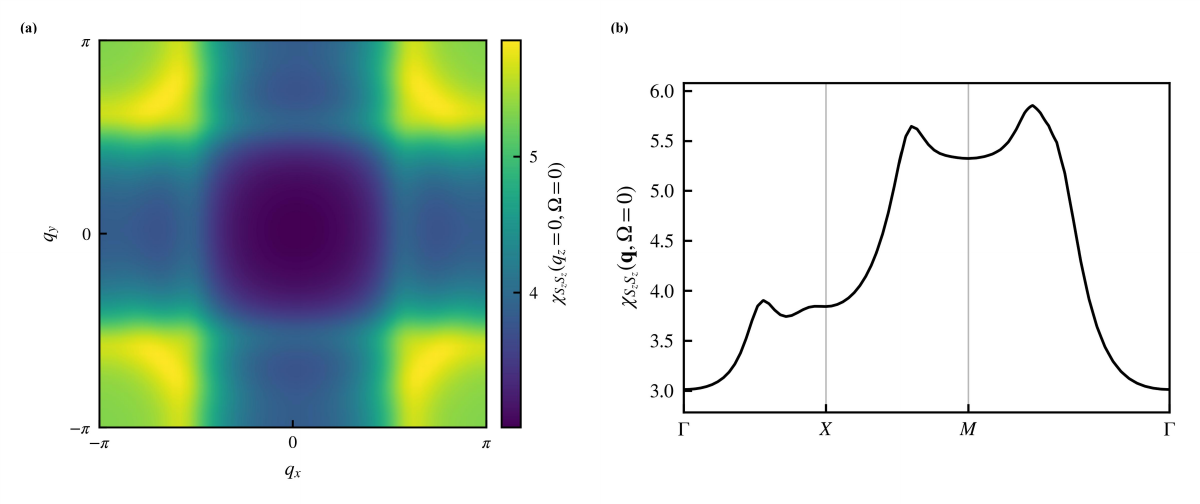}
\caption{(Colour online) DBSE spin susceptibility of the correlated normal state.  (a) Static
  $\chi^{\rm DBSE}_{S_zS_z}(\mathbf q,q_z=0,\Omega=0)$ in the two-dimensional Brillouin zone,
  calculated with $\beta=25~{\rm eV}^{-1}$ and $N_{\rm wf}=10$.  (b) Corresponding line cut
  along $\Gamma$--$X$--$M$--$\Gamma$.  The susceptibility is weak near $\Gamma$ and enhanced
  at finite momenta close to the Brillouin-zone boundary, demonstrating that the correlated
  spin response remains dominated by antiferromagnetic or near-antiferromagnetic
  fluctuations after local two-particle vertex corrections are included.}
\label{fig:dbse_susceptibility}
\end{figure*}

Fig.~\ref{fig:dbse_susceptibility}(a) shows that the DBSE spin susceptibility is strongly
momentum dependent.  The response is minimal around $\Gamma$ and is enhanced at finite
momenta, with broad maxima near the zone boundary and around the $M$ region.  The line cut
in Fig.~\ref{fig:dbse_susceptibility}(b) makes the same point more explicitly: the dominant
magnetic fluctuations are not ferromagnetic $q\simeq0$ fluctuations, but finite-$q$ spin
fluctuations capable of connecting different portions of the bilayer Fermi surface.  This
is precisely the magnetic environment required for repulsive spin fluctuations to generate a
sign-changing singlet order parameter.

The momentum structure of the DBSE response reinforces the mechanism identified from the
pocket-pair decomposition in Fig.~\ref{fig:pocket_decomposition}.  The correlated magnetic
spectrum is broad rather than concentrated at a single sharp nesting vector, which is
favorable to the distributed inter-pocket scattering network that stabilizes the
$A_{1g}$ $s_\pm$ state.  In contrast, the bare-RPA $B_{2g}$ state relies more strongly on a
narrow set of $\gamma$-centered scattering processes that are selectively weakened by the
orbital-dependent DMFT self-energy.  The DBSE result therefore supports the central physical
interpretation: strong correlations reshape the spin response while preserving robust
finite-momentum magnetic fluctuations, converting the weak-coupling $d_{xy}$ tendency into
a correlation-stabilized $s_\pm$ hierarchy.

Thus, the DBSE susceptibility provides a vertex-level confirmation that the magnetic
background used in the self-energy-renormalized RPA analysis is not an artifact of the
unrenormalized RPA ladder.  The persistence of broad finite-$q$ spin fluctuations strengthens
the result that a reliable prediction of the pairing symmetry in \LNO\ requires the
correlated quasiparticle and vertex-renormalized magnetic response of the strongly
interacting normal state.

\section{Discussion and conclusions}
\label{sec:discussion}

We have shown that the leading superconducting instability of the four-orbital bilayer
Wannier model is qualitatively changed when the RPA particle-hole bubble is dressed by a
single-site DMFT self-energy.  In the ordinary RPA benchmark the dominant eigenvalue lies
in the $B_{2g}$ $d_{xy}$ channel, consistent with the tendency of weak-coupling treatments
to place bilayer nickelates close to a boundary between sign-changing $s$ wave and $d$ wave.
After replacing $G_0G_0$ by $G_{\rm DMFT}G_{\rm DMFT}$, the $A_{1g}$ $s_\pm$ state becomes
leading, the $B_{2g}$ state is strongly suppressed, and the remaining $d$-wave competitor
has mainly $B_{1g}$ character.  This reversal reflects the orbital selectivity of the
correlated normal state: the DMFT self-energy filters the $d_{3z^2-r^2}$- and
$\gamma$-sheet-rich processes that support the bare-RPA $d_{xy}$ state, while the $A_{1g}$ solution survives
through a broader network of inter-pocket scatterings, especially the $\alpha\beta$ and
$\beta\gamma$ channels.  The DBSE susceptibility provides a two-particle consistency check:
a robust finite-momentum spin response remains after the local DMFT particle-hole vertex is
included, supporting the spin-fluctuation background behind the $A_{1g}$ solution.

These results support a picture in which strong correlations do not simply suppress
spin-fluctuation pairing in \LNO.  Instead, the orbital-dependent self-energy removes a
weakness of ordinary RPA--its tendency to overemphasize coherent bare-Fermi-surface
scattering that favors $d_{xy}$ pairing--and stabilizes a sign-changing $A_{1g}$ state
compatible with the experimentally observed nodeless or U-shaped gap phenomenology.  The
prediction of the pairing symmetry in \LNO\ therefore requires more than an accurate
Wannier Hamiltonian; it also requires a correlated quasiparticle propagator appropriate to
the strongly renormalized normal state.  Our results identify strong electronic correlations
as an essential ingredient for a reliable prediction of the superconducting gap symmetry in
pressurized La$_3$Ni$_2$O$_7$.

\medskip
\noindent\begin{minipage}{\columnwidth}
\textit{Data Availability Statement.}\quad
The data and code that support the findings of this article are available from the corresponding
author upon reasonable request.
\end{minipage}\par

\bibliography{references}

\end{document}